\documentclass[sigconf,authorversion]{acmart}

\usepackage[utf8]{inputenc}
\usepackage[english]{babel}
\usepackage{blindtext}
\usepackage{acronym}
\usepackage{listings}
\usepackage{hhline}
\usepackage{units}
\usepackage{balance}

\copyrightyear{2018}
\acmYear{2018}
\setcopyright{acmlicensed}
\acmConference[EPIQ'18]{Workshop on the Evolution, Performance, and Interoperability of QUIC }{December 4, 2018}{Heraklion, Greece}
\acmBooktitle{Workshop on the Evolution, Performance, and Interoperability of QUIC (EPIQ'18), December 4, 2018, Heraklion, Greece}
\acmPrice{15.00}
\acmDOI{10.1145/3284850.3284853}
\acmISBN{978-1-4503-6082-1/18/12}


\newacro{symex}[SymEx]{Symbolic Execution}

\newcommand{\afblock}[1]{\noindent{\textbf{#1}}}
\newcommand{\emblock}[1]{\noindent{\emph{#1}}}

\begin{document}
\title{Interoperability-Guided Testing of QUIC \\Implementations using Symbolic Execution}

\author{Felix Rath, Daniel Schemmel, Klaus Wehrle}
\affiliation{%
  \department{Communication and Distributed Systems}
  \institution{RWTH Aachen University}
  \city{Aachen}
  \country{Germany}
}
\email{{firstname.lastname}@comsys.rwth-aachen.de}

\begin{abstract}
  The main reason for the standardization of network protocols, like QUIC, is to ensure interoperability between implementations, which poses a challenging task.
  Manual tests are currently used to test the different existing implementations for interoperability, but given the complex nature of network protocols, it is hard to cover all possible edge cases.

  State-of-the-art automated software testing techniques, such as \ac{symex}, have proven themselves capable of analyzing complex real-world software and finding hard to detect bugs.
  We present a \ac{symex}-based method for finding interoperability issues in QUIC implementations, and explore its merit in a case study that analyzes the interoperability of picoquic and QUANT.

  We find that, while \ac{symex} is able to analyze deep interactions between different implementations and uncovers several bugs, in order to enable efficient interoperability testing, implementations need to provide additional information about their current protocol state.

\end{abstract}

\maketitle

\acresetall{}

\section{Introduction}
The emergence of new, modern protocols for the Internet promises a solution to long-standing issues that can only be solved by changing core parts of the current protocol stack.
Such new protocols and their implementations must meet the highest requirements: They will have to reliably function at similar levels of maturity as what they aim to replace.
This includes aspects such as reliability, security, performance and, prominently, interoperability between implementations.

Ensuring interoperability is the main reason for standardizing QUIC as a protocol, and the IETF standardization process goes to great lengths, such as requiring multiple independent implementations, to make sure this is achievable.
Thus, better methods and tools that assist with the difficult challenge of interoperability testing are highly desirable.

Automated testing techniques, such as \ac{symex}, have proven themselves to be capable of analyzing complex real world software, usually focused on finding low-level safety violations~\cite{ThreeDecadesLater2013}, and \ac{symex} has also proven its worth in the networking domain in various other ways~\cite{KleeNet2008,KleeNet2010,SOFT2012,PIC2015,SymbexNet2014,SymNet2016,SymPerf2017,VigNAT2017}.

This paper explores the potential of \ac{symex} for checking the interoperability of QUIC implementations.
It does so by presenting a \ac{symex}-based method to detect interoperability issues, and demonstrates its potential in a case study of two existing QUIC implementations, picoquic and QUANT.
We discover that, while our method is able to successfully analyze nontrivial interactions between different implementations, implementations need to disclose more protocol-level information to truly enable deep semantic interoperability testing.

\subsection{Key Contributions and Outline}
The key contributions of this paper are as follows:

\begin{itemize}
  \item We describe a method that uses \acf{symex} to test QUIC implementations for interoperability, and discuss how additional information from implementations about their current protocol state could be leveraged for semantically deeper testing.
  \item We then present our case study in which we symbolically test picoquic and QUANT for interoperability, and discuss the abstraction layers that are necessary to enable \ac{symex} for QUIC implementations.
  \item The final key contribution is the evaluation of our implementation, testing picoquic and QUANT, in which we report on the performance of our method as well as on defects we discovered.
\end{itemize}

\begin{figure*}[t]
  \includegraphics[width=\textwidth]{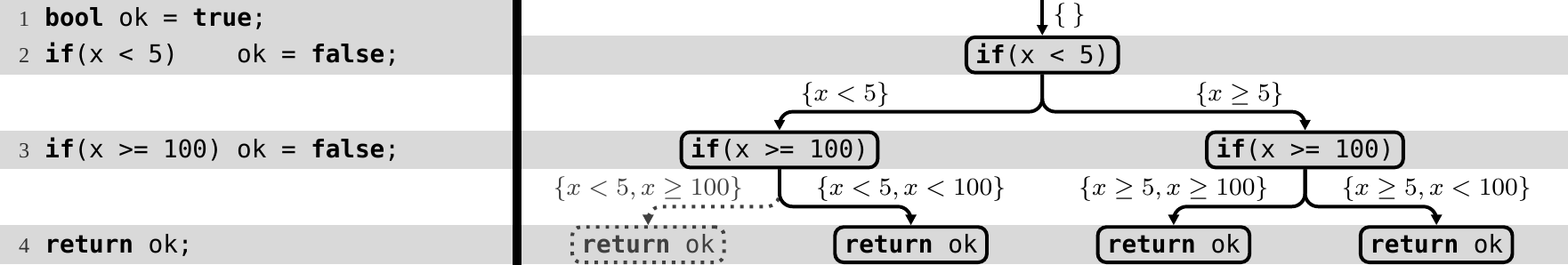}
  \caption{\acf{symex} of a small example program. Constraints encountered in branching statements are recorded in the path constraints of the corresponding explored paths. By checking new branching conditions for satisfiability on each path, exactly all reachable paths through the program are explored.}
  \label{fig:symex}
\end{figure*}

We begin by giving background on \ac{symex} in Sect.~\ref{sec:symex}, followed by a discussion of related work in Sect.~\ref{sec:rw}. We then present our method in Sect.~\ref{sec:method}, and describe its implementation and the setup of the case study in Sect.~\ref{sec:casestudy}. This is followed by an evaluation of our results in Sect.~\ref{sec:eval}, before we shortly discuss future work in Sect.~\ref{sec:fw} and conclude in Sect.~\ref{sec:conclusion}.

\section{Symbolic Execution (SymEx)}
\label{sec:symex}

Given a program that takes some input (e.g., command line arguments, files, network packets, etc.), \ac{symex} systematically explores the program by executing all reachable paths.
It does so by assigning symbolic values instead of concrete ones to its input, which allows the \ac{symex} engine to fork execution at a branch-statement (i.e., \texttt{if}) when both branches are feasible.
If this is the case, the condition that caused the fork (i.e., the condition inside the \texttt{if} statement) is remembered on the execution path following the \texttt{true}-branch as an additional constraint.
On the other execution path, which follows the \texttt{false}-branch, the negation of the condition is remembered as a constraint instead.
To determine the reachability given the current constraints, an SMT solver, such as Z3~\cite{Z3SMT2008}, is queried.
SMT solvers are the backbone of every \ac{symex} engine, and their performance and completeness directly influence the efficiency of the symbolic analysis, and they ensure that only feasible paths are explored.

Continuing in this fashion a \ac{symex} engine will explore all reachable paths through the program.
Whenever a path terminates, either regularly or because an error was encountered, the engine will query the SMT solver using the collected path constraints to get concrete values for each symbolic input value.
These values will then be recorded in the form of a test case, which can then be run again later to exercise the same path through the program.
If a bug was encountered, the generated test case will be able to reproduce the taken path for further debugging and introspection.

Figure~\ref{fig:symex} shows a small example program that performs operations depending on the value of a symbolic input variable \texttt{x}.
The program contains two conditional branches that have to be traversed before the \texttt{return} in line 4 is reached.
On the right, all paths explored by \ac{symex} are shown.

In the beginning, \texttt{x} is unconstrained, but, as \ac{symex} progresses, a path for each side of the first branch is explored.
For each side, a corresponding constraint (either $x<5$ or $x\geq5$) is added to the path constraints.
When the second branch is reached, only three paths need to be explored further: The constraint set $\{x<5,x\geq100\}$ is not satisfiable, and therefore this path will never be reachable during execution.
In the end, \ac{symex} will query the SMT solver for concrete values for \texttt{x} for each path to generate a suite of concrete test cases that cover all reachable paths of the program.

\section{Related Work}
\label{sec:rw}
Formal methods have long been used to analyze network protocols~\cite{FormalProtocolTesting1989,DivideAbstractModelCheck1999,SSL30Analysis1998,SPINCrypto2002,CMC2002}, often with a focus on security.
However, even if the formal analysis of a network protocol has successfully proven a property, be it related to correctness or security, it is by no means guaranteed that this property will also hold for an implementation of said protocol.

Programs have also been analyzed with formal methods, such as \ac{symex}, to test for obvious problems like memory safety and assertion violations~\cite{KLEE2008,ThreeDecadesLater2013} and for less easily checked properties, such as liveness violations~\cite{SymLive2018} and authentication bypass flaws in firmware binaries~\cite{Firmalice2015}.
One of the main problems encountered when formally analyzing real-world code is the penchant of the state-space to grow infeasibly large---a problem also known as \emph{state explosion}.
Many different approaches to tame the state explosion problem inherent in \ac{symex} have been proposed in the past: state merging~\cite{EfficientStateMerging2012}, targeted search strategies~\cite{DirectedSymEx2011} and pruning of provably equivalent paths~\cite{RWset2008}, to name a few.

As the the state explosion problem grows exponentially with the number of programs considered at the same time, approaches explicitly targeting distributed programs have been developed.
For example, KleeNet~\cite{KleeNet2008,KleeNet2010} exploits the independence of networked programs by delaying codependent path forks until messages are received at each node that require the fork to be actualized.

Testing protocols and programs independently is -- however worthwhile -- not enough.
To this end, approaches have been designed that test implementations for protocol compliance using many different testing and verification methodologies, ranging from fuzzing~\cite{SNOOZE2006,MutationFuzz2012} over \ac{symex}~\cite{SymbexNet2014} to model checking~\cite{ModelCheckingProtocolImplementations2004,CMC2002}.
Validating that a given implementation fulfills a specification or standard does, however, require a formalized representation to be available, which effectively constitutes another implementation of the specification.

One way to circumvent this chicken-egg problem is to exploit the fact that any relevant standard will have multiple implementations, which enables the substitution of compliance testing with that of interoperability testing.
While it is possible that neither implementation is technically compliant with the standard, it becomes more and more improbable that the standard is captured incorrectly by many different people in exactly the same manner.
Due to the inherent state-explosion problem of interoperability testing (multiple different, or even all possible programs are considered at once), multiple approaches to specialized~\cite{SOFT2012} and general~\cite{PIC2015, SymbexNet2014} interoperability testing have been proposed in the past.

\section{Method Outline}
\label{sec:method}
\ac{symex} engines such as KLEE~\cite{KLEE2008}, which our cases study utilizes, usually expect their input to be a program.
However, protocol implementations are naturally libraries, and as such lack an implicit singular entry point.
Although ways to analyze libraries directly have been proposed~\cite{UCKLEE2015}, they suffer from a lack of insight into what constitutes a \emph{sensible} use of the library.
Instead, we propose to analyze programs that utilize the libraries and execute different test scenarios.
One way to choose the test scenarios for this would be to use existing applications that already implement real-world application logic.
This is currently difficult to do for QUIC, as there are only very few applications built on top of QUIC, and is further complicated by their use of only a small set of different QUIC implementations.

Instead, we follow the current best-practice in compliance testing by designing test scenarios based on primitives defined in the QUIC standard.
Unlike common, concrete compliance testing suites, we formulate symbolic testing scenarios that perform large families of related tests in one go.
These describe the involved endpoints (e.g., clients and servers) and the communication that takes place between them, for example, which connections are established, which streams are opened, what is sent on those streams, and so on.
Such scenarios can be defined in both high-level as well as low-level terms.
A more low-level scenario describes individual packets and effects such as loss or reordering instead of focusing on connections and streams.


Independently of the test scenarios, we need to define what we categorize as actual \emph{errors}, so that the \ac{symex} engine can actually detect which paths exhibit erroneous behavior.
We present two categories of errors here, one focused on interoperability, and one focused on robustness.

\subsection{Testing Interoperability}
Generally speaking, whenever there is a conflict between what the communication partners believe the state of their connection to be, an interoperability violation exists.
In the case of networked programs, it is important to quantify the belief state of each endpoint in a way that is neither too constrained (e.g., if the server believes that a data connection is open, but the client has already sent a shutdown request, there is no conflict), nor too open (otherwise error detection becomes impossible).
Such issues can cause the communication to continue without exhibiting low-level errors, but the result of the execution to differ from what was expected.
For example, if the amount of application data sent by one endpoint differs from the amount of data received by the other after a finished transmission, this is an error, as the two endpoints hold different beliefs about the correct state of the connection.

To be able to detect such bugs, it is necessary to have a way to extract the current belief state of each endpoint in a way that can be compared to that of the other endpoints.
Here, standardization can help:
A definition of what exactly is part of the (belief) state of a QUIC connection could be used by implementations to provide this information to analysis tools.
Such information could then be used by testing and verification tools to great effect, enabling stronger and more semantically meaningful analyses.

\subsection{Testing Robustness}
Robustness can be defined as the ability of an implementation to deal correctly with unexpected events, such as packet loss, reordering or packets crafted with malicious intent.
Here, errors usually manifest in the form of, e.g., out-of-bound memory accesses, use-after-free violations, assertion errors, etc.
When using a \ac{symex} engine, the engine will provide the capability to test for such violations out-of-the-box, already providing valuable testing feedback without needing to define additional error conditions.

\subsection{Generality of the Method}
This method is, in its core, protocol independent, and can be applied to other protocols than QUIC.
However, its application to QUIC shows the effort required to implement it for non-trivial, real world protocols, as well as its suitability for such protocols.
The question in this case is scalability: While it is usually straight-forward to apply any method to simple examples, we are interested in whether the method scales to implementations of complex protocols, such as QUIC, and also to see the requirements for such protocols in regards to automated testing.

\section{Case Study: Picoquic and QUANT}
\label{sec:casestudy}
For our case study we implemented our method for picoquic\footnote{\url{https://github.com/private-octopus/picoquic}} and QUANT\footnote{\url{https://github.com/NTAP/quant}}.
These implementations were chosen because they are written in C, and could therefore be analyzed by KLEE, our \ac{symex} engine of choice, out-of-the-box.
It has successfully been shown that \ac{symex} can also be applied to programs written in other languages, like C++~\cite{KLOVER2011}, so this is not a limitation of the general approach.

We defined multiple test scenarios, and developed simple clients for each library that execute the defined scenarios. An additional challenge is that the KLEE \ac{symex} engine~\cite{KLEE2008} only works on single programs, which caused us to implement a single program that instantiates all communication partners and advances them in tandem for each test scenario.
This means that one endpoint (e.g., client or server) is executed until it can make no more progress (i.e., it blocks waiting for a response) at which point execution switches to the next endpoint.
This continues until either the scenario is finished, or one endpoint reports an error.

In the following sections we describe our test scenarios and present which library-specific adaptations were necessary, as well as which library-independent abstractions we implemented that can be re-used for other libraries in the future.


\subsection{Test Scenarios}
We decided upon three test scenarios that exercise some of the core features of QUIC:
In the first scenario, a client establishes a connection with a server, then closes it again.
In the second scenario, we establish a connection just as before, but the client also opens a stream and sends a simple HTTP request (\texttt{GET /index.html}), which the server then closes without responding.
Finally, the third scenario builds upon the second one, but the server also responds with a one-byte response.
We define as interoperability issues any case in which a run ends without the underlying scenario being fulfilled, e.g., because a connection could not be established, or because one of the endpoints timed out during the process.

For each library we implemented a frontend which provides functions that create a client or a server for one of the scenarios, a function that advances a client or server (executing it until it has reached the next stage in the scenario or is blocked on network input), and a function that checks whether a client or server is finished with the scenario.
In our evaluation, we focused on the scenarios being executed with a picoquic client communicating with a QUANT server, but our implementation also supports the other cases (QUANT client and picoquic server or both from the same library).

\begin{figure}[t]
  \includegraphics[]{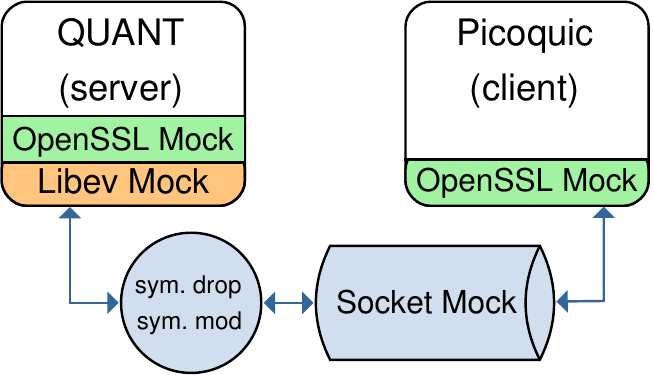}
  \caption{Case study setup: A QUANT server and a picoquic client communicate via a channel that symbolically drops and modifies packets.}
  \label{fig:system}
\end{figure}

\subsection{Library-Independent Abstractions}
\label{sec:abstractions}
In order symbolically execute the test scenarios, we had to implement abstractions for various functionalities, such as blocking and non-blocking network operations, as well as cryptographic operations.
Figure~\ref{fig:system} shows an overview of the test setup, including the layers we replaced with abstractions, such as communication via UDP.

\afblock{UNIX Sockets.} To enable KLEE to correctly route network data, we explicitly modeled the network environment by providing simple custom implementations of functions such as \texttt{socket}, \texttt{connect}, \texttt{sendto}, and so forth.

Note that we implemented only those parts of the POSIX socket API necessary for executing the two implementations, as the whole API surface covers extensive functionality.
These parts where straightforward to implement, and we used a simple linked-list structure for sent packets and otherwise tracked additional information per socket, and only implemented UDP functionality.


\emblock{Symbolic Values.} We also used our abstraction to model some of the properties of UDP-based communication, such as unreliability.
To model packet drops, we used a symbolic variable that decides whether or not to drop each packet.
The result of this is that we will for each packet explore a path in which this packet was lost.

Additionally, we also implemented the possibility to make certain bytes of a sent packet symbolic instead of simply delivering the packet.
This allows testing the receiver of each packet with regards to robustness, within the current state of the communication.

While this is enough for QUIC implementations that rely on blocking communication, such as picoquic, others rely on asynchronous event notifications.
For the case of QUANT, this is provided by libev.

\afblock{Libev.} Libev is a library that provides an event loop for asynchronous applications.
We implemented a mock version of libev that fulfilled our requirements of being easy to integrate into QUANT and our final test scenario binaries, as well as being simple to evaluate with KLEE.

\afblock{OpenSSL.} \ac{symex} is not able to reverse constraints that are based on cryptographic operations (encryption, decryption, hashing, etc.), as otherwise the underlying cryptography would be broken.
As QUIC heavily relies on cryptographic operations, we needed to make these operations transparent, for which we implemented an OpenSSL abstraction that always performs null-encryption.
This means we implemented most of the functions used by picoquic and QUANT in a very bare-bones fashion, often nothing more than a no-op, and implemented encryption and decryption basically as a \texttt{memcpy}.

With regards to hash-functions, we decided to use actual implementations of these hashes instead of, e.g., hashing all values to the same hash value.
This had certain implications for our evaluation: On one hand, it makes our implementation more correct, as different messages will correctly hash to different values.
On the other hand, whenever our \ac{symex} engine had to reverse the result of such a hash-function, it would not be able to do so, possibly preventing the exploration of certain parts of the libraries.

\afblock{Library-Independence.} All of these mocks are implemented independently of the QUIC library under test, and are reusable for future interoperability tests.
Thus, our work lays the foundation for testing a larger set of implementations.

\subsection{Picoquic}
For picoquic, a frontend that can execute our test scenarios was straightforward to implement, due to the fact that an example client and an example server were available.
We replaced blocking reads in the client and server with points at which execution would return to the test harness, so the next communication partner would be able to make progress.

This was made easy by the fact that the picoquic API itself only prepares packets for sending, and leaves the actual sending to the application.
This means that we could implement the communication handover inside of our frontend library, and did not need to implement it inside of picoquic itself, requiring no changes to the library.

\begin{table*}[t]
\begin{tabular}{|c||r|c|c|c|c|c|c|}
\hline
Configuration & Instrs/s & Time[h] & ICov[\%] & BCov[\%] & TSolver[\%] & MaxMem[GB] & Unique errors \\ \hhline{|=#=|=|=|=|=|=|=|}
sym-stream    & \texttt{1725742}  & 0:01    & 38.96    & 24.81    & 0.06        & 0.16       & 2             \\ \hline
sym-version   & \texttt{232139}   & 0:25    & 38.87    & 24.83    & 83.83       & 0.15       & 1             \\ \hline
sym-drop      & \texttt{432753}   & 8:00    & 38.85    & 25.31    & 0.02        & 11.97      & 1             \\ \hline
sym-mod-1     & \texttt{380751}   & 8:00    & 41.10    & 27.04    & 0.79        & 32.44      & 0             \\ \hline
sym-mod-5     & \texttt{241116}   & 7:00    & 40.11    & 26.11    & 8.35        & 33.02      & 0             \\ \hline
sym-mod-10    & \texttt{4118}     & 8:01    & 32.11    & 18.79    & 78.78       & 5.34       & 1             \\ \hline
\end{tabular}
\caption{Results of our evaluation after running each configuration with a time limit of 8 hours.
  Together all configurations discovered 5 unique errors. Instrs/s: Executed instructions per second. Time[h]: Time after which the execution terminated. ICov[\%]: Achieved instruction coverage in percent. BCov[\%]: Achieved branch coverage in percent. TSolver[\%]: Time of the execution that was spent solving SMT queries. MaxMem[GB]: Maximum memory usage in GB. Unique errors: Errors reported that were not reported by other runs.}
\label{fig:table}
\end{table*}

\subsection{QUANT}
The changes needed for QUANT were more extensive, as QUANT internally uses a libev-based event loop, which we needed to intercept in order to be able to return execution to the test harness when the event loop would block waiting for new data.
To do so, in addition to implementing a simple variant of libev as described before, we modified the top-level API functions of QUANT.
These expected blocking behavior of the underlying event loop, but instead we changed them to return control to the test harness when entering the event loop would block.

Additionally, QUANT made use of global variables, which lead to corrupt behavior when, e.g., trying to instantiate a QUANT server and client in the same binary.
To circumvent this, we performed a simple renaming of all defined symbols on the LLVM IR of QUANT.
This prefixes all functions, such as \texttt{q\_connect}, with a prefix of our choice, resulting in, e.g., \texttt{client\_\_q\_connect} and \texttt{server\_\_q\_connect}.
Since this also renamed all global variables, this allowed us to test QUANT clients and servers in the same binary.

As only a single QUANT instance is contained, our case study did not require this additional renaming.
However, since global variables are a common feature in programs, it is necessary that this is also supported by our approach.

\section{Evaluation}
\label{sec:eval}
For our evaluation we considered six different combinations of scenarios and symbolic input.
All configurations were executed in KLEE with Z3~\cite{Z3SMT2008} as the underlying SMT solver, with a time limit of 8 hours and a memory limit of \unit[32]{GB} on a system with two E5-2643 v4 processors providing a total of 12 physical cores and \unit[256]{GB} main memory.
We additionally added timeouts of 10 seconds per instruction and per query, to prevent the analysis from being stuck on too hard queries.

We chose picoquic and QUANT for our case study as both are written in C, which is supported by KLEE.
Both of these libraries also implement the newest version of the QUIC standard at the time of writing (draft 14).

\subsection {Configurations}
We tested six configurations and provide the results of their symbolic execution in Table~\ref{fig:table}.

\afblock{Sym-stream.} This configuration combines all three described scenarios.
We added symbolic input that chooses which of these to execute, resulting in the execution of all three, as \ac{symex} explores all possible paths.
This is the same as executing all three scenarios concretely without further symbolic values.
This configuration terminated in about a minute after exploring all three reachable paths through the test binary, and reported two bugs.

The first error is an interoperability bug that we originally found during the development of our implementation.
This bug occurs in the second scenario, when the stream is closed by the server without sending any data.
In this case, the QUANT server silently closes the stream on its end, not notifying the client.
The client then times out and closes the connection prematurely.

The second error occurs because certain resources are freed which might still be in use inside of libev.
This bug was discovered because our libev-abstraction touched the freed value, which was discovered by KLEE.
In practice, this kind of bug is hard to check: As it occurred in a shared library, concrete execution with a tool like ASAN would not detect this bug, and this is exactly the kind of bug that can cause rare, random crashes.
We reported both bugs in the QUANT library, and both were verified and later fixed\footnote{\url{https://github.com/NTAP/quant/issues/16}}\footnote{\url{https://github.com/NTAP/quant/issues/17}}.

This configuration also gives a good baseline for instruction and branch coverage, as the other configurations explore the third scenario, which covers the most API surface, with different symbolic values.
The values for instruction and branch coverage include dead code, and thus their absolute values need to be treated with care.
However, they can be used to compare against the other configurations.

\afblock{Sym-version.} This configuration is built upon the third scenario (connection establishment, new stream, response), but makes the version proposed by the picoquic client to the server symbolic.
We chose this configuration because setting the proposed version is an option of the picoquic library.

This configuration terminates after only 25 minutes, with most of the time spent inside the solver.
The reason for this is that this configuration resulted early on in the generation of constraints that were not solvable by the SMT solver in the given timeout, thus terminating all paths early on.
Nevertheless, this configuration also found an error that prevented the establishment of a connection.

This error occurs when the proposed version is set to \texttt{0xbabababa}.
As the current QUIC draft reserves all versions of the form \texttt{0x?a?a?a?a} for version negotiation, it seems plausible that this version by itself could not lead to a successfully established connection.
We categorize this bug as very mild, as it is obviously only a small API problem.

\afblock{Sym-drop.} For this configuration we symbolically dropped every packet.
This is the first configuration that needed more than a few hundred MB of memory, and also the first that explored a large number of paths through the program.
We see that only little time was spent solving constraints, which makes sense, since no symbolic data was actually touched by either of the libraries (either a packet was delivered as-is or it was dropped).


Most interestingly, this is also the first configuration that found a bug which only occurred after multiple exchanged packets, and would not be easily found during manual testing.
The reported test case drops the 4th, 5th and 7th packet exchanged between the two endpoints, which triggers a segfault due to a null-pointer in the QUANT server when the 9th packet is received.

We verified that this bug also occurs when running concretely with regular OpenSSL instead of our abstraction.
This is a robustness bug, but it might also be an interoperability bug, as other implementations might not trigger it.

\afblock{Sym-mod-X.} In these configurations, the first X bytes of every sent packet are made symbolic, in order to test the robustness of the receiving endpoint.
This category includes the two configurations that reached the highest instruction and branch coverages, but it also includes the configuration that achieved the lowest coverages.
A trend can be seen here: More symbolic bytes cause more work for the \ac{symex} engine due to state explosion, resulting in more time spent inside the SMT solver, resulting in slower progress overall.

However, the run that achieved the lowest coverage uncovered an additional bug in QUANT's packet receiving code.
The generated test case triggers the bug by replacing the first 10 bytes of the first packet sent by picoquic with the concrete values \texttt{[0xff, 0x01, 0x01, 0x01, 0x01, 0x67, 0xff, 0xff, 0xff, 0xff]}, which leads to a null-pointer dereference in the server.
We verified this bug as well while running without our OpenSSL abstraction.

\balance
\section{Future Work}
\label{sec:fw}
While our case study shows the usefulness of automated testing techniques such as \ac{symex} for analyzing QUIC implementations, there is still much that can be done.
A first and important step is the definition of the kinds of belief state QUIC implementations should be able to report on.
In a second step, such a model can then be used for testing implementations for state divergence regarding the belief states of the different endpoints. 
Most of the effort to achieve this should be in defining a common ground for the definition of the belief state.
We expect then extracting the belief state from implementations to require manageable effort, since implementations must already be keeping track of the state of each connection.

This could be realized in the form of standardized testing and verification interfaces for protocol implementations, which would enable high levels of accessibility for new analysis approaches.
This in turn would lead to high-quality implementations, increasing stability, robustness and performance in a field where all of these are important.

Our test scenario only dropped packets or made some of the bytes symbolic, but did not take the specific structure of QUIC packets into account.
Here, a layer that reads the packets that are sent and performs symbolic mutations based on the semantics of the protocol, e.g., symbolic ACK numbers, could lead to more thorough and scalable testing.
The need for such a method becomes obvious when looking at the \texttt{sym-mod-10} configuration, which already caused a visible slowdown of the \ac{symex} engine due to state explosion.

Furthermore, to analyze more parts of protocol implementations, additional test scenarios that excercise so-far uncovered protocol functionality are required.
One way to achieve this would be to create more test scenarios based on the QUIC standard.
However, it might also be possible to automatically derive test scenarios, either from a model of the standard, or from the implementations themselves.
For this, knowing which API call caused which state change could help choosing possible next API calls.

To extend test scenarios to more than two endpoints, it might be favorable to utilize \ac{symex} techniques that target distributed systems, such as KleeNet~\cite{KleeNet2008,KleeNet2010}.
While doing so, it might also become relevant to investigate \emph{symbolic time}, since behavior in network protocols is often dependent on timing, most notably due to timeouts.

\section{Conclusion}
\label{sec:conclusion}
We presented an interoperability-guided method to test QUIC implementations and demonstrated its potential in a case study.
Our method consists of testing implementations in pre-defined scenarios, but enriched with additional symbolic input, such as packet drops and symbolic modifications.

In our case study we showed that, in order to symbolically execute and test implementations, it is required that underlying libraries are abstracted in a way that is sensible for testing.
On one hand, kernel code such as UNIX sockets can otherwise not be executed and analyzed, but also to, e.g., turn encryption transparent in order to enable any analysis at all.

We were able to uncover several bugs with varying levels of severity.
While two were simple API issues that could easily be found through manual testing, two of the other three would be hard to find without some kind of automated testing approach, as they occur only in very specific situations: The right packets in a long chain of packets have to be dropped, or a very specific first packet has to be sent.
The last bug was only detected due to abstracting libev, but does not necessarily require \ac{symex} to uncover.

In summary, most of these bugs are robustness bugs.
To detect deeper semantic interoperability bugs, support in implementations that provides information about the current belief state of endpoints is required.
We appeal to the authors of QUIC implementations, as well as to the members of the IETF working group, to develop a common understanding of what information makes up the belief state of a QUIC connection, and to extend implementations with ways to report this information for the sake of deep semantic interoperability testing.

\section*{Acknowledgements}
This research is supported by the European Research Council (ERC) under the European Union's Horizon 2020 Research and Innovation Programme (grant agreement \textnumero{}.~647295 (SYMBIOSYS)).

\bibliographystyle{ACM-Reference-Format}
\bibliography{reference}


\begin{thebibliography}{26}


\ifx \showCODEN    \undefined \def \showCODEN     #1{\unskip}     \fi
\ifx \showDOI      \undefined \def \showDOI       #1{#1}\fi
\ifx \showISBNx    \undefined \def \showISBNx     #1{\unskip}     \fi
\ifx \showISBNxiii \undefined \def \showISBNxiii  #1{\unskip}     \fi
\ifx \showISSN     \undefined \def \showISSN      #1{\unskip}     \fi
\ifx \showLCCN     \undefined \def \showLCCN      #1{\unskip}     \fi
\ifx \shownote     \undefined \def \shownote      #1{#1}          \fi
\ifx \showarticletitle \undefined \def \showarticletitle #1{#1}   \fi
\ifx \showURL      \undefined \def \showURL       {\relax}        \fi
\providecommand\bibfield[2]{#2}
\providecommand\bibinfo[2]{#2}
\providecommand\natexlab[1]{#1}
\providecommand\showeprint[2][]{arXiv:#2}

\bibitem[\protect\citeauthoryear{Banks, Cova, Felmetsger, Almeroth, Kemmerer,
  and Vigna}{Banks et~al\mbox{.}}{2006}]%
        {SNOOZE2006}
\bibfield{author}{\bibinfo{person}{Greg Banks}, \bibinfo{person}{Marco Cova},
  \bibinfo{person}{Viktoria Felmetsger}, \bibinfo{person}{Kevin Almeroth},
  \bibinfo{person}{Richard Kemmerer}, {and} \bibinfo{person}{Giovanni Vigna}.}
  \bibinfo{year}{2006}\natexlab{}.
\newblock \showarticletitle{{{SNOOZE}}: {{Toward}} a {{Stateful NetwOrk
  prOtocol fuzZEr}}}. In \bibinfo{booktitle}{{\em Information {{Security}}}}
  {\em (\bibinfo{series}{Lecture Notes in Computer Science})},
  \bibfield{editor}{\bibinfo{person}{Sokratis~K. Katsikas},
  \bibinfo{person}{Javier L\'opez}, \bibinfo{person}{Michael Backes},
  \bibinfo{person}{Stefanos Gritzalis}, {and} \bibinfo{person}{Bart Preneel}}
  (Eds.). \bibinfo{publisher}{{Springer Berlin Heidelberg}},
  \bibinfo{pages}{343--358}.
\newblock
\showISBNx{978-3-540-38343-7}
\showDOI{%
\url{https://doi.org/10.1007/11836810_25}}


\bibitem[\protect\citeauthoryear{Boonstoppel, Cadar, and Engler}{Boonstoppel
  et~al\mbox{.}}{2008}]%
        {RWset2008}
\bibfield{author}{\bibinfo{person}{Peter Boonstoppel},
  \bibinfo{person}{Cristian Cadar}, {and} \bibinfo{person}{Dawson Engler}.}
  \bibinfo{year}{2008}\natexlab{}.
\newblock \showarticletitle{{{RWset}}: {{Attacking Path Explosion}} in
  {{Constraint}}-{{Based Test Generation}}}. In \bibinfo{booktitle}{{\em Tools
  and {{Algorithms}} for the {{Construction}} and {{Analysis}} of {{Systems}}}}
  {\em (\bibinfo{series}{Lecture Notes in Computer Science})},
  \bibfield{editor}{\bibinfo{person}{C.~R. Ramakrishnan} {and}
  \bibinfo{person}{Jakob Rehof}} (Eds.). \bibinfo{publisher}{{Springer Berlin
  Heidelberg}}, \bibinfo{pages}{351--366}.
\newblock
\showISBNx{978-3-540-78800-3}
\showDOI{%
\url{https://doi.org/10.1007/978-3-540-78800-3_27}}


\bibitem[\protect\citeauthoryear{Cadar, Dunbar, and Engler}{Cadar
  et~al\mbox{.}}{2008}]%
        {KLEE2008}
\bibfield{author}{\bibinfo{person}{Cristian Cadar}, \bibinfo{person}{Daniel
  Dunbar}, {and} \bibinfo{person}{Dawson~R. Engler}.}
  \bibinfo{year}{2008}\natexlab{}.
\newblock \showarticletitle{{{KLEE}}: {{Unassisted}} and {{Automatic
  Generation}} of {{High}}-{{Coverage Tests}} for {{Complex Systems
  Programs}}.}. In \bibinfo{booktitle}{{\em Proceedings of the 8th {{USENIX
  Symposium}} on {{Operating Systems Design}} and {{Implementation}}
  ({{OSDI}}'08)}}, Vol.~\bibinfo{volume}{8}. \bibinfo{pages}{209--224}.
\newblock


\bibitem[\protect\citeauthoryear{Cadar and Sen}{Cadar and Sen}{2013}]%
        {ThreeDecadesLater2013}
\bibfield{author}{\bibinfo{person}{Cristian Cadar} {and}
  \bibinfo{person}{Koushik Sen}.} \bibinfo{year}{2013}\natexlab{}.
\newblock \showarticletitle{Symbolic {{Execution}} for {{Software Testing}}:
  {{Three Decades Later}}}.
\newblock \bibinfo{journal}{{\em Commun. ACM\/}} \bibinfo{volume}{56},
  \bibinfo{number}{2} (\bibinfo{date}{Feb.} \bibinfo{year}{2013}),
  \bibinfo{pages}{82--90}.
\newblock
\showISSN{0001-0782}
\showDOI{%
\url{https://doi.org/10.1145/2408776.2408795}}


\bibitem[\protect\citeauthoryear{{de Moura} and Bj\o{}rner}{{de Moura} and
  Bj\o{}rner}{2008}]%
        {Z3SMT2008}
\bibfield{author}{\bibinfo{person}{Leonardo {de Moura}} {and}
  \bibinfo{person}{Nikolaj Bj\o{}rner}.} \bibinfo{year}{2008}\natexlab{}.
\newblock \showarticletitle{Z3: {{An Efficient SMT Solver}}}. In
  \bibinfo{booktitle}{{\em Tools and {{Algorithms}} for the {{Construction}}
  and {{Analysis}} of {{Systems}}}} {\em (\bibinfo{series}{Lecture Notes in
  Computer Science})}, \bibfield{editor}{\bibinfo{person}{C.~R. Ramakrishnan}
  {and} \bibinfo{person}{Jakob Rehof}} (Eds.). \bibinfo{publisher}{{Springer
  Berlin Heidelberg}}, \bibinfo{pages}{337--340}.
\newblock
\showISBNx{978-3-540-78800-3}
\showDOI{%
\url{https://doi.org/10.1007/978-3-540-78800-3_24}}


\bibitem[\protect\citeauthoryear{Kuznetsov, Kinder, Bucur, and
  Candea}{Kuznetsov et~al\mbox{.}}{2012}]%
        {EfficientStateMerging2012}
\bibfield{author}{\bibinfo{person}{Volodymyr Kuznetsov},
  \bibinfo{person}{Johannes Kinder}, \bibinfo{person}{Stefan Bucur}, {and}
  \bibinfo{person}{George Candea}.} \bibinfo{year}{2012}\natexlab{}.
\newblock \showarticletitle{Efficient {{State Merging}} in {{Symbolic
  Execution}}}. In \bibinfo{booktitle}{{\em Proceedings of the 33rd {{ACM
  SIGPLAN Conference}} on {{Programming Language Design}} and
  {{Implementation}}}} {\em (\bibinfo{series}{PLDI '12})}.
  \bibinfo{publisher}{{ACM}}, \bibinfo{address}{New York, NY, USA},
  \bibinfo{pages}{193--204}.
\newblock
\showISBNx{978-1-4503-1205-9}
\showDOI{%
\url{https://doi.org/10.1145/2254064.2254088}}


\bibitem[\protect\citeauthoryear{Ku\'zniar, Pere{\v s}\'ini, Canini, Venzano,
  and Kosti\'c}{Ku\'zniar et~al\mbox{.}}{2012}]%
        {SOFT2012}
\bibfield{author}{\bibinfo{person}{Maciej Ku\'zniar}, \bibinfo{person}{Peter
  Pere{\v s}\'ini}, \bibinfo{person}{Marco Canini}, \bibinfo{person}{Daniele
  Venzano}, {and} \bibinfo{person}{Dejan Kosti\'c}.}
  \bibinfo{year}{2012}\natexlab{}.
\newblock \showarticletitle{A {{SOFT Way}} for {{Openflow Switch
  Interoperability Testing}}}. In \bibinfo{booktitle}{{\em Proceedings of the
  8th {{International Conference}} on {{Emerging Networking Experiments}} and
  {{Technologies}}}} {\em (\bibinfo{series}{CoNEXT '12})}.
  \bibinfo{publisher}{{ACM}}, \bibinfo{address}{New York, NY, USA},
  \bibinfo{pages}{265--276}.
\newblock
\showISBNx{978-1-4503-1775-7}
\showDOI{%
\url{https://doi.org/10.1145/2413176.2413207}}


\bibitem[\protect\citeauthoryear{Li, Ghosh, and Rajan}{Li
  et~al\mbox{.}}{2011}]%
        {KLOVER2011}
\bibfield{author}{\bibinfo{person}{Guodong Li}, \bibinfo{person}{Indradeep
  Ghosh}, {and} \bibinfo{person}{Sreeranga~P. Rajan}.}
  \bibinfo{year}{2011}\natexlab{}.
\newblock \showarticletitle{{{KLOVER}}: {{A Symbolic Execution}} and
  {{Automatic Test Generation Tool}} for {{C}}++ {{Programs}}}. In
  \bibinfo{booktitle}{{\em Computer {{Aided Verification}}}} {\em
  (\bibinfo{series}{Lecture Notes in Computer Science})},
  \bibfield{editor}{\bibinfo{person}{Ganesh Gopalakrishnan} {and}
  \bibinfo{person}{Shaz Qadeer}} (Eds.). \bibinfo{publisher}{{Springer Berlin
  Heidelberg}}, \bibinfo{pages}{609--615}.
\newblock
\showISBNx{978-3-642-22110-1}


\bibitem[\protect\citeauthoryear{Ma, Yit~Phang, Foster, and Hicks}{Ma
  et~al\mbox{.}}{2011}]%
        {DirectedSymEx2011}
\bibfield{author}{\bibinfo{person}{Kin-Keung Ma}, \bibinfo{person}{Khoo
  Yit~Phang}, \bibinfo{person}{Jeffrey~S. Foster}, {and}
  \bibinfo{person}{Michael Hicks}.} \bibinfo{year}{2011}\natexlab{}.
\newblock \showarticletitle{Directed {{Symbolic Execution}}}. In
  \bibinfo{booktitle}{{\em Static {{Analysis}}}} {\em (\bibinfo{series}{Lecture
  Notes in Computer Science})}, \bibfield{editor}{\bibinfo{person}{Eran Yahav}}
  (Ed.). \bibinfo{publisher}{{Springer Berlin Heidelberg}},
  \bibinfo{pages}{95--111}.
\newblock
\showISBNx{978-3-642-23702-7}
\showDOI{%
\url{https://doi.org/10.1007/978-3-642-23702-7_11}}


\bibitem[\protect\citeauthoryear{Maggi and Sisto}{Maggi and Sisto}{2002}]%
        {SPINCrypto2002}
\bibfield{author}{\bibinfo{person}{Paolo Maggi} {and} \bibinfo{person}{Riccardo
  Sisto}.} \bibinfo{year}{2002}\natexlab{}.
\newblock \showarticletitle{Using {{SPIN}} to {{Verify Security Properties}} of
  {{Cryptographic Protocols}}}. In \bibinfo{booktitle}{{\em Model {{Checking
  Software}}}} {\em (\bibinfo{series}{Lecture Notes in Computer Science})},
  \bibfield{editor}{\bibinfo{person}{Dragan Bo{\v s}na{\v c}ki} {and}
  \bibinfo{person}{Stefan Leue}} (Eds.). \bibinfo{publisher}{{Springer Berlin
  Heidelberg}}, \bibinfo{pages}{187--204}.
\newblock
\showISBNx{978-3-540-46017-6}
\showDOI{%
\url{https://doi.org/10.1007/3-540-46017-9_14}}


\bibitem[\protect\citeauthoryear{Mitchell, Shmatikov, and Stern}{Mitchell
  et~al\mbox{.}}{1998}]%
        {SSL30Analysis1998}
\bibfield{author}{\bibinfo{person}{John~C. Mitchell}, \bibinfo{person}{Vitaly
  Shmatikov}, {and} \bibinfo{person}{Ulrich Stern}.}
  \bibinfo{year}{1998}\natexlab{}.
\newblock \showarticletitle{Finite-State {{Analysis}} of {{SSL}} 3.0}. In
  \bibinfo{booktitle}{{\em Proceedings of the 7th {{USENIX Security
  Symposium}}}} {\em (\bibinfo{series}{SSYM'98})}. \bibinfo{publisher}{{USENIX
  Association}}, \bibinfo{address}{Berkeley, CA, USA}, \bibinfo{pages}{16--16}.
\newblock


\bibitem[\protect\citeauthoryear{Musuvathi and Engler}{Musuvathi and
  Engler}{2004}]%
        {ModelCheckingProtocolImplementations2004}
\bibfield{author}{\bibinfo{person}{Madanlal Musuvathi} {and}
  \bibinfo{person}{Dawson~R. Engler}.} \bibinfo{year}{2004}\natexlab{}.
\newblock \showarticletitle{Model {{Checking Large Network Protocol
  Implementations}}}. In \bibinfo{booktitle}{{\em {{NSDI}} '04}}.
  \bibinfo{pages}{155--168}.
\newblock


\bibitem[\protect\citeauthoryear{Musuvathi, Park, Chou, Engler, and
  Dill}{Musuvathi et~al\mbox{.}}{2002}]%
        {CMC2002}
\bibfield{author}{\bibinfo{person}{Madanlal Musuvathi}, \bibinfo{person}{David
  Y.~W. Park}, \bibinfo{person}{Andy Chou}, \bibinfo{person}{Dawson~R. Engler},
  {and} \bibinfo{person}{David~L. Dill}.} \bibinfo{year}{2002}\natexlab{}.
\newblock \showarticletitle{{{CMC}}: {{A Pragmatic Approach}} to {{Model
  Checking Real Code}}}. In \bibinfo{booktitle}{{\em Proceedings of the 5th
  {{Symposium}} on {{Operating Systems Design}} and {{implementationCopyright
  Restrictions Prevent ACM}} from {{Being Able}} to {{Make}} the {{PDFs}} for
  {{This Conference Available}} for {{Downloading}}}} {\em
  (\bibinfo{series}{OSDI '02})}. \bibinfo{publisher}{{USENIX Association}},
  \bibinfo{address}{Berkeley, CA, USA}, \bibinfo{pages}{75--88}.
\newblock
\showISBNx{978-1-4503-0111-4}


\bibitem[\protect\citeauthoryear{Pedrosa, Fogel, Kothari, Govindan, Mahajan,
  and Millstein}{Pedrosa et~al\mbox{.}}{2015}]%
        {PIC2015}
\bibfield{author}{\bibinfo{person}{Luis Pedrosa}, \bibinfo{person}{Ari Fogel},
  \bibinfo{person}{Nupur Kothari}, \bibinfo{person}{Ramesh Govindan},
  \bibinfo{person}{Ratul Mahajan}, {and} \bibinfo{person}{Todd Millstein}.}
  \bibinfo{year}{2015}\natexlab{}.
\newblock \showarticletitle{Analyzing {{Protocol Implementations}} for
  {{Interoperability}}}. In \bibinfo{booktitle}{{\em Proceedings of the 12th
  {{USENIX Conference}} on {{Networked Systems Design}} and
  {{Implementation}}}} {\em (\bibinfo{series}{NSDI'15})}.
  \bibinfo{publisher}{{USENIX Association}}, \bibinfo{address}{Berkeley, CA,
  USA}, \bibinfo{pages}{485--498}.
\newblock
\showISBNx{978-1-931971-21-8}


\bibitem[\protect\citeauthoryear{Ramos and Engler}{Ramos and Engler}{2015}]%
        {UCKLEE2015}
\bibfield{author}{\bibinfo{person}{David~A. Ramos} {and}
  \bibinfo{person}{Dawson Engler}.} \bibinfo{year}{2015}\natexlab{}.
\newblock \showarticletitle{Under-{{Constrained Symbolic Execution}}:
  {{Correctness Checking}} for {{Real Code}}}. In \bibinfo{booktitle}{{\em
  Proceedings of the 24th {{USENIX Conference}} on {{Security Symposium}}}}
  {\em (\bibinfo{series}{SEC'15})}. \bibinfo{publisher}{{USENIX Association}},
  \bibinfo{address}{Berkeley, CA, USA}, \bibinfo{pages}{49--64}.
\newblock
\showISBNx{978-1-931971-23-2}


\bibitem[\protect\citeauthoryear{Rath, Krude, R\"uth, Schemmel, Hohlfeld, Link,
  and Wehrle}{Rath et~al\mbox{.}}{2017}]%
        {SymPerf2017}
\bibfield{author}{\bibinfo{person}{Felix Rath}, \bibinfo{person}{Johannes
  Krude}, \bibinfo{person}{Jan R\"uth}, \bibinfo{person}{Daniel Schemmel},
  \bibinfo{person}{Oliver Hohlfeld}, \bibinfo{person}{J\'o \'Agila~Bitsch
  Link}, {and} \bibinfo{person}{Klaus Wehrle}.}
  \bibinfo{year}{2017}\natexlab{}.
\newblock \showarticletitle{{{SymPerf}}: {{Predicting Network Function
  Performance}}}. In \bibinfo{booktitle}{{\em Proceedings of the {{SIGCOMM
  Posters}} and {{Demos}}}} {\em (\bibinfo{series}{SIGCOMM Posters and Demos
  '17})}. \bibinfo{publisher}{{ACM}}, \bibinfo{address}{New York, NY, USA},
  \bibinfo{pages}{34--36}.
\newblock
\showISBNx{978-1-4503-5057-0}
\showDOI{%
\url{https://doi.org/10.1145/3123878.3131977}}


\bibitem[\protect\citeauthoryear{Sasnauskas, Landsiedel, Alizai, Weise,
  Kowalewski, and Wehrle}{Sasnauskas et~al\mbox{.}}{2010}]%
        {KleeNet2010}
\bibfield{author}{\bibinfo{person}{Raimondas Sasnauskas}, \bibinfo{person}{Olaf
  Landsiedel}, \bibinfo{person}{Muhammad~Hamad Alizai},
  \bibinfo{person}{Carsten Weise}, \bibinfo{person}{Stefan Kowalewski}, {and}
  \bibinfo{person}{Klaus Wehrle}.} \bibinfo{year}{2010}\natexlab{}.
\newblock \showarticletitle{{{KleeNet}}: {{Discovering Insidious Interaction
  Bugs}} in {{Wireless Sensor Networks Before Deployment}}}. In
  \bibinfo{booktitle}{{\em Proceedings of the 9th {{ACM}}/{{IEEE International
  Conference}} on {{Information Processing}} in {{Sensor Networks}}}} {\em
  (\bibinfo{series}{IPSN '10})}. \bibinfo{publisher}{{ACM}},
  \bibinfo{address}{New York, NY, USA}, \bibinfo{pages}{186--196}.
\newblock
\showISBNx{978-1-60558-988-6}
\showDOI{%
\url{https://doi.org/10.1145/1791212.1791235}}


\bibitem[\protect\citeauthoryear{Sasnauskas, Link, Alizai, and
  Wehrle}{Sasnauskas et~al\mbox{.}}{2008}]%
        {KleeNet2008}
\bibfield{author}{\bibinfo{person}{Raimondas Sasnauskas}, \bibinfo{person}{J\'o
  \'Agila~Bitsch Link}, \bibinfo{person}{Muhammad~Hamad Alizai}, {and}
  \bibinfo{person}{Klaus Wehrle}.} \bibinfo{year}{2008}\natexlab{}.
\newblock \showarticletitle{{{KleeNet}}: {{Automatic Bug Hunting}} in {{Sensor
  Network Applications}}}. In \bibinfo{booktitle}{{\em Proceedings of the 6th
  {{ACM Conference}} on {{Embedded Network Sensor Systems}}}} {\em
  (\bibinfo{series}{SenSys '08})}. \bibinfo{publisher}{{ACM}},
  \bibinfo{address}{New York, NY, USA}, \bibinfo{pages}{425--426}.
\newblock
\showISBNx{978-1-59593-990-6}
\showDOI{%
\url{https://doi.org/10.1145/1460412.1460485}}


\bibitem[\protect\citeauthoryear{Schemmel, B\"uning, Soria~Dustmann, Noll, and
  Wehrle}{Schemmel et~al\mbox{.}}{2018}]%
        {SymLive2018}
\bibfield{author}{\bibinfo{person}{Daniel Schemmel}, \bibinfo{person}{Julian
  B\"uning}, \bibinfo{person}{Oscar Soria~Dustmann}, \bibinfo{person}{Thomas
  Noll}, {and} \bibinfo{person}{Klaus Wehrle}.}
  \bibinfo{year}{2018}\natexlab{}.
\newblock \showarticletitle{Symbolic {{Liveness Analysis}} of {{Real}}-{{World
  Software}}}. In \bibinfo{booktitle}{{\em Computer {{Aided Verification}}}}
  {\em (\bibinfo{series}{Lecture Notes in Computer Science})},
  \bibfield{editor}{\bibinfo{person}{Hana Chockler} {and}
  \bibinfo{person}{Georg Weissenbacher}} (Eds.). \bibinfo{publisher}{{Springer
  International Publishing}}, \bibinfo{pages}{447--466}.
\newblock
\showISBNx{978-3-319-96142-2}
\showDOI{%
\url{https://doi.org/10.1007/978-3-319-96142-2_27}}


\bibitem[\protect\citeauthoryear{Shoshitaishvili, Wang, Hauser, Kruegel, and
  Vigna}{Shoshitaishvili et~al\mbox{.}}{2015}]%
        {Firmalice2015}
\bibfield{author}{\bibinfo{person}{Yan Shoshitaishvili}, \bibinfo{person}{Ruoyu
  Wang}, \bibinfo{person}{Christophe Hauser}, \bibinfo{person}{Christopher
  Kruegel}, {and} \bibinfo{person}{Giovanni Vigna}.}
  \bibinfo{year}{2015}\natexlab{}.
\newblock \showarticletitle{Firmalice - {{Automatic Detection}} of
  {{Authentication Bypass Vulnerabilities}} in {{Binary Firmware}}}. In
  \bibinfo{booktitle}{{\em Proceedings of the 2015 {{Network}} and
  {{Distributed System Security}} ({{NDSS}}'15)}}.
\newblock
\showDOI{%
\url{https://doi.org/10.14722/ndss.2015.23294}}


\bibitem[\protect\citeauthoryear{Sidhu and Leung}{Sidhu and Leung}{1989}]%
        {FormalProtocolTesting1989}
\bibfield{author}{\bibinfo{person}{Deepinder~P. Sidhu} {and}
  \bibinfo{person}{Ting-Kau Leung}.} \bibinfo{year}{1989}\natexlab{}.
\newblock \showarticletitle{Formal Methods for Protocol Testing: A Detailed
  Study}.
\newblock \bibinfo{journal}{{\em IEEE Transactions on Software Engineering\/}}
  \bibinfo{volume}{15}, \bibinfo{number}{4} (\bibinfo{date}{April}
  \bibinfo{year}{1989}), \bibinfo{pages}{413--426}.
\newblock
\showISSN{0098-5589}
\showDOI{%
\url{https://doi.org/10.1109/32.16602}}


\bibitem[\protect\citeauthoryear{Song, Cadar, and Pietzuch}{Song
  et~al\mbox{.}}{2014}]%
        {SymbexNet2014}
\bibfield{author}{\bibinfo{person}{JaeSeung Song}, \bibinfo{person}{Cristian
  Cadar}, {and} \bibinfo{person}{Peter Pietzuch}.}
  \bibinfo{year}{2014}\natexlab{}.
\newblock \showarticletitle{{{SymbexNet}}: {{Testing Network Protocol
  Implementations}} with {{Symbolic Execution}} and {{Rule}}-{{Based
  Specifications}}}.
\newblock \bibinfo{journal}{{\em IEEE Transactions on Software Engineering\/}}
  \bibinfo{volume}{40}, \bibinfo{number}{7} (\bibinfo{date}{July}
  \bibinfo{year}{2014}), \bibinfo{pages}{695--709}.
\newblock
\showISSN{0098-5589}
\showDOI{%
\url{https://doi.org/10.1109/TSE.2014.2323977}}


\bibitem[\protect\citeauthoryear{Stahl, Baukus, Lakhnech, and Stefen}{Stahl
  et~al\mbox{.}}{1999}]%
        {DivideAbstractModelCheck1999}
\bibfield{author}{\bibinfo{person}{Karsten Stahl}, \bibinfo{person}{Kai
  Baukus}, \bibinfo{person}{Yassine Lakhnech}, {and} \bibinfo{person}{Martin
  Stefen}.} \bibinfo{year}{1999}\natexlab{}.
\newblock \showarticletitle{Divide, {{Abstract}}, and {{Model}}-{{Check}}}. In
  \bibinfo{booktitle}{{\em Theoretical and {{Practical Aspects}} of {{SPIN
  Model Checking}}}} {\em (\bibinfo{series}{Lecture Notes in Computer
  Science})}, \bibfield{editor}{\bibinfo{person}{Dennis Dams},
  \bibinfo{person}{Rob Gerth}, \bibinfo{person}{Stefan Leue}, {and}
  \bibinfo{person}{Mieke Massink}} (Eds.). \bibinfo{publisher}{{Springer Berlin
  Heidelberg}}, \bibinfo{pages}{57--76}.
\newblock
\showISBNx{978-3-540-48234-5}


\bibitem[\protect\citeauthoryear{Stoenescu, Popovici, Negreanu, and
  Raiciu}{Stoenescu et~al\mbox{.}}{2016}]%
        {SymNet2016}
\bibfield{author}{\bibinfo{person}{Radu Stoenescu}, \bibinfo{person}{Matei
  Popovici}, \bibinfo{person}{Lorina Negreanu}, {and} \bibinfo{person}{Costin
  Raiciu}.} \bibinfo{year}{2016}\natexlab{}.
\newblock \showarticletitle{{{SymNet}}: {{Scalable Symbolic Execution}} for
  {{Modern Networks}}}. In \bibinfo{booktitle}{{\em Proceedings of the 2016
  {{ACM SIGCOMM Conference}}}} {\em (\bibinfo{series}{SIGCOMM '16})}.
  \bibinfo{publisher}{{ACM}}, \bibinfo{address}{New York, NY, USA},
  \bibinfo{pages}{314--327}.
\newblock
\showISBNx{978-1-4503-4193-6}
\showDOI{%
\url{https://doi.org/10.1145/2934872.2934881}}


\bibitem[\protect\citeauthoryear{Zaostrovnykh, Pirelli, Pedrosa, Argyraki, and
  Candea}{Zaostrovnykh et~al\mbox{.}}{2017}]%
        {VigNAT2017}
\bibfield{author}{\bibinfo{person}{Arseniy Zaostrovnykh},
  \bibinfo{person}{Solal Pirelli}, \bibinfo{person}{Luis Pedrosa},
  \bibinfo{person}{Katerina Argyraki}, {and} \bibinfo{person}{George Candea}.}
  \bibinfo{year}{2017}\natexlab{}.
\newblock \showarticletitle{A {{Formally Verified NAT}}}. In
  \bibinfo{booktitle}{{\em Proceedings of the {{Conference}} of the {{ACM
  Special Interest Group}} on {{Data Communication}}}} {\em
  (\bibinfo{series}{SIGCOMM '17})}. \bibinfo{publisher}{{ACM}},
  \bibinfo{address}{New York, NY, USA}, \bibinfo{pages}{141--154}.
\newblock
\showISBNx{978-1-4503-4653-5}
\showDOI{%
\url{https://doi.org/10.1145/3098822.3098833}}


\bibitem[\protect\citeauthoryear{Zhang, Wen, and Tang}{Zhang
  et~al\mbox{.}}{2012}]%
        {MutationFuzz2012}
\bibfield{author}{\bibinfo{person}{Zhao Zhang}, \bibinfo{person}{Qiao-Yan Wen},
  {and} \bibinfo{person}{Wen Tang}.} \bibinfo{year}{2012}\natexlab{}.
\newblock \showarticletitle{An {{Efficient Mutation}}-{{Based Fuzz Testing
  Approach}} for {{Detecting Flaws}} of {{Network Protocol}}}. In
  \bibinfo{booktitle}{{\em 2012 {{International Conference}} on {{Computer
  Science}} and {{Service System}}}}. \bibinfo{pages}{814--817}.
\newblock
\showDOI{%
\url{https://doi.org/10.1109/CSSS.2012.208}}


\end{thebibliography}

\end{document}